\documentclass[twocolumn,traditabstract,longauth]{aa}
\usepackage{amssymb}
\usepackage{mathptmx}
\usepackage{natbib}
\usepackage[]{graphicx}
\defcitealias{canameras15}{C15}
\defcitealias{canameras18a}{C18}

\begin{document}

\title{Planck's Dusty GEMS. VII. Atomic carbon {\bf and molecular gas} in dusty starburst galaxies at $z=2$ to $4$
  \thanks{Based on observations obtained with the 30-m telescope and
    the Plateau de Bure interferometer of IRAM under program IDs
    082-12, D05-12, D09-12, 094-13, 223-13, 108-14, and 217-14}}
  \author{N.~P.~H.~Nesvadba\inst{1,2},
    R.~Canameras\inst{3}, R.~Kneissl\inst{4,5}, S.~Koenig\inst{6},
C.~Yang\inst{4}, E.~Le~Floc'h\inst{7}, A.~Omont\inst{8}, D.~Scott\inst{9}}
\institute{Institut d'Astrophysique Spatiale, CNRS, Universite
Paris-Sud, Universite Paris-Saclay, 91405 Orsay, France \and email:
nicole.nesvadba@ias.u-psud.fr \and Dark Cosmology Centre, Niels Bohr
Institute, University of Copenhagen, Juliane Maries Vej 30, DK-2100
Copenhagen, Denmark \and European Southern Observatory, ESO Vitacura,
Alonso de Cordova 3107, Vitacura, Casilla 19001 Santiago, Chile \and
Atacama Large Millimeter/submillimeter Array, ALMA Santiago Central
Offices, Alonso de Cordova 3107, Vitacura, Cailla 763-0355, Santiago,
Chile, \and Chalmers University of Technology, Onsala Space Observatory,
Onsala, Sweden, 
\and Laboratoire AIM, CEA/DSM/IRFU, CNRS, Universite
Paris-Diderot, Bat 709, 91191 Gif-sur-Yvette, France \and Institut
d'Astrophysique de Paris, UPMC Universite Paris 06, UMR 7095, 75014
Paris, France \and Department of Physics and Astronomy, University of
British Columbia, 6224 Agricultural Road, Vancouver, 6658 British
Columbia, Canada}

\titlerunning{Planck's Dusty Gems: Atomic gas probed through [CI]}

\authorrunning{Nesvadba et al.}  

\date{Received / Accepted }
\abstract{
The bright $^3$P$_1$-$^3$P$_0$ ([CI] 1--0) and $^3$P$_2$-$^3$P$_1$
([CI] 2--1) lines of atomic carbon are becoming more and more widely
employed tracers of the cold neutral gas in high-redshift galaxies.
Here we present observations of these lines in the 11 galaxies of the
set of {\it Planck}'s Dusty GEMS, the brightest gravitationally lensed
galaxies on the extragalactic submillimeter sky probed by the {\it
Planck} satellite. We have [CI] 1--0 measurements for seven, and [CI]
2--1 measurements for eight galaxies, including four galaxies where
both lines are measured. We use our observations to constrain the gas
excitation mechanism, excitation temperatures, optical depths, 
atomic carbon and molecular gas masses, and carbon
abundances. Ratios of $L_{\rm CI}/L_{\rm FIR}$ are similar to those found in
the local Universe, and suggest that the total cooling budget through
atomic carbon has not strongly changed in the last 12~Gyr. Both
lines are optically thin and trace $1\ -\ 6 \times 10^7$
M$_{\odot}$ of atomic carbon. Carbon abundances, $X_{\rm CI}$, are between 2.5
and $4\times 10^{-5}$, for a ''ULIRG'' CO-to-H$_2$
conversion factor of $\alpha_{C\rm O}=0.8\ {\rm M}_{\odot}$ / [K km s$^{-1}$
pc$^2$] .  Ratios of molecular gas masses derived from [CI] 1--0 and
CO agree within the measurement uncertainties for five galaxies, and
to better than a factor of~2 for another two with [CI] 1--0 measurements,
after taking CO excitation carefully into account. This does not
support the idea that intense, high-redshift starburst galaxies host large quantities
of ''CO-dark'' gas. These results support the common assumptions
underlying most molecular gas mass estimates made for massive, dusty,
high-redshift starburst galaxies, although the good agreement between
the masses obtained with both tracers cannot be taken as an
independent confirmation of either $\alpha_{\rm CO}$ or $X_{\rm CI}$.}

\keywords{galaxies: high-redshift}

\maketitle
\section{Introduction}
\label{sec:introduction}

Numerous observations in recent years have firmly established that the
vigorous star-formation episodes in massive, dusty starburst galaxies
at redshifts $z\ge2$, which form most of the stellar populations in
these galaxies within a few hundred Myr, are fueled by massive
reservoirs of dense molecular gas \citep[e.g.][see \citealt{solomon05}
and \citealt{carilli13} for reviews]{tacconi08, ivison11,
riechers13}. The physical and kinematic properties of this gas, such
as densities and mass surface densities, temperatures, and bulk and
turbulent motion, are critical for understanding the regulation and
upper limits imposed on the vigorous star formation up to the highest
star-formation rates found in the Universe.

Thanks to the new generation of wide-band millimeter and
sub-millimeter receivers, and sensitive interferometers like ALMA and
IRAM's NOEMA, we are now able to study these gaseous reservoirs in
galaxies in the early Universe at an interesting level of detail,
extending and complementing the classical CO emission-line studies
through observations of additional tracers. This includes the
fine-structure lines of atomic or singly ionized carbon, [CI], and
[CII], which are excellent tracers of the cold neutral gas in
galaxies, or various other tracers of denser gas. \citet{beuther16}
argue that [CI] is the best tracer of the cold neutral medium, because
[CII] can also be associated with ionized gas, whereas CO emission
only probes fairly dense molecular gas, and misses more diffuse gas
that is however seen in [CI]. \citet{goldsmith12} even argue that [CI]
emission can be associated with low-excitation gas seen in [CII]
absorption, as also found observationally in the Milky Way
\citep[][]{gerin15} and at high redshift in the Garnet
\citep[PLCK$\_$G045.1$+$61.1, ][]{nesvadba16}, a source whose [CI]
properties we will also discuss here. [CI], [CII], and CO are
therefore complementary probes of the gas in high-redshift galaxies.

\begin{table*}
\caption{Targets and details of our [CI] observations. We list the
source name, right ascension and declination, redshift, luminosity
distance, observed far-infrared luminosity, transition, tuning
frequency, date of our observations, time spent on the target,
and root mean square of the resulting spectrum. Dots indicate
lines outside of the atmospheric windows \label{tab:observations}.}

\centering
\begin{tabular}{lcccccccccc}
\hline
\hline
Source      &  RA          & Dec        & Redshift & $D_{\rm L}$   & $ \mu_{\rm gas}\ L_{\rm FIR}$      &  Trans. & $\nu_0$   & date         & ToT & rms   \\ 
            &  (J2000)       &  (J2000)   &          & [Gpc]& [$10^{13} L_{\odot}$]  &       & [GHz]    & [mm/dd/yy] & [min] & [mK]  \\
\hline
PLCK$\_$G045.1+61.1 & 15:02:36.04  & $+$29:20:51& 3.43    & 29.86 & 8.4$\pm$0.1        & 1--0   & 111.035  & 02/03/14   & 81    & 1.17   \\
PLCK$\_$G045.1+61.1 &              &            &          &       &                    & 2--1    & \dots   & \dots        & \dots & \dots  \\
\hline
PLCK$\_$G080.2+49.8 & 15:44:32.40 & $+$50:23:46 & 2.60    & 21.79 &  4.6$\pm$0.1       & 1--0   & 136.749  & 02/03/14   & 122   & 2.35   \\
PLCK$\_$G080.2+49.8 &             &             &          &       &                    & 2--1   & 224.400  & 02/03/14   & 162   & 0.82   \\
\hline
PLCK$\_$G092.5+42.9 & 16:09:17.76 & $+$60:45:21 & 3.26    & 28.61 & 24.8$\pm$02        & 1--0   & 115.639  & 04/19/14 \& 23/02/15   & 314   & 1.25   \\
PLCK$\_$G092.5+42.9 &             &             &          &       &                    & 2--1   & \dots    & \dots        & \dots & \dots  \\
\hline 
PLCK$\_$G102.1+53.6 & 14:29:17.98 & $+$51:29:09 & 2.92    & 24.99 & 7.9$\pm$0.1        & 1--0   & \dots    & \dots        & \dots & \dots  \\
PLCK$\_$G102.1+53.6 &             &             &          &       &                    & 2--1   & 206.267  & 02/19,21,23/15   & 81 & 0.5    \\
\hline
PLCK$\_$G113.7+61.0 & 13:23:02.88 & $+$55:36:01 & 2.41    & 19.88 & 9.9$\pm$0.2        & 1--0   &  144.160 & 02/21/15   & 202   & 0.3    \\
PLCK$\_$G113.7+61.0 &             &             &          &       &                    & 2--1   &  236.666 & 02/19/15   & 150   & 1.2    \\
\hline
PLCK$\_$G138.6+62.0 & 12:02:07.68 & $+$53:34:40 & 2.44    & 20.20 & 9.0$\pm$0.1        & 1--0   & 143.677  & 07/06/13   & 80    & 0.8    \\
PLCK$\_$G138.6+62.0 &             &             &          &       &                    & 2--1   & 231.300  & 02/19/15   & 102   & 2.0    \\
\hline
PLCK$\_$G145.2+50.9 & 10:53:2.56 & $+$60:51:49  & 3.55    & 31.76 & 21.8$\pm$0.2       & 1--0   & 108.167  & 05/06/14   & 120   & 1.3     \\
PLCK$\_$G145.2+50.9 &            &              &          &       &                    & 2--1   & \dots    & \dots        & \dots & \dots  \\
\hline
PLCK$\_$G165.7+67.0 & 11:27:14.60 & $+$42:28:25 & 2.24    & 18.18 & 10.3$\pm$0.1       & 1--0   & 152.070  & 01/31/14,02/01-04/14 & 172 & 0.4   \\
PLCK$\_$G165.7+67.0 &             &             &          &       &                    & 2--1   & 245.500  & 02/19-20/15 &  126       & 0.7       \\
\hline
PLCK$\_$G200.6+46.1 & 09:32:23.67 & $+$27:25:00 & 2.97    & 25.14 & 5.7$\pm$0.1        & 1--0   & \dots    & \dots      & \dots   & \dots  \\
PLCK$\_$G200.6+46.1 &             &             &          &       &                    & 2--1   &  206.276 & 02/21/15 & 162      & 1.3     \\
\hline
PLCK$\_$G231.3+72.2 & 11:39:21.60 & $+$20:24:53 & 2.86    & 24.00 & 7.5$\pm$0.1        & 1--0   & \dots    & \dots      & \dots   & \dots  \\
PLCK$\_$G231.3+72.2 &             &             &          &       &                    & 2--1   & 209.100  & 02/20/15 & 120     & 0.911  \\
\hline
PLCK$\_$G244.8+54.9 & 10:53:53.04 & $+$05:56:21 & 3.01    & 25.47 & 26.5$\pm$0.2       & 1--0   & \dots    & \dots      & \dots   & \dots  \\
PLCK$\_$G244.8+54.9 &             &             &          &       &                    & 2--1   & 205.200  & 02/19/15 & 120     & 2.08   \\
\hline
\hline
\end{tabular}
\end{table*}

Atomic carbon, specifically, is probed through two bright transitions,
$^3$P$_1$-$^3$P$_0$, ([CI] 1--0), and $^3$P$_2$-$^3$P$_1$ ([CI] 2--1)
at rest-frame frequencies of 492.1607~GHz and 809.3435~GHz,
respectively, which are conveniently redshifted into the millimeter
and lower sub-millimeter regime for redshifts $z\sim2-4$. With upper
level energies of $E_{{\rm up},10}=24.2$~K and $E_{{\rm
up},21}=62.5$~K, and critical densities of about $10^3$ cm$^{-3}$,
they are easily excited over large ranges in gas density and
temperature, from fairly diffuse gas \citep[][]{philips81, gerin00,
goldsmith12} to gas within dense molecular clouds
\citep[][]{papadopoulos04a}. This makes them useful global probes of
the cold neutral medium in very high-redshift galaxies. Perhaps most
importantly, both [CI] lines remain optically thin even in very dusty,
vigorous starburst galaxies, which is a clear advantage for mass
estimates ensuring that mass scales linearly with line
luminosity. However, other systematic uncertainties remain, e.g.,
related to the carbon abundance, with uncertainties of factors of a
few \citep[e.g.,][]{alaghband13}. Observations of CO, in contrast, are
notoriously plagued by uncertainties related to most of the gas being
hidden within optically thick clouds. This adds considerable
uncertainty when generalizing the results of these observations to the
overall molecular gas component in galaxies, without falling back on
empirical relationships whose use cannot always be justified easily 
for each individual galaxy and analysis.

The [CI] fine structure lines arise also from physically relatively
simple systems, so that many of their physical characteristics can be
calculated directly from their observed properties and measured
brightness temperatures or line fluxes, e.g., masses and abundances,
or their contribution to the cooling budget. Several studies suggest
also that they are less sensitive than CO to variations in metallicity
\citep[which can lead to significant reservoirs of so-called
``CO-dark'' gas, e.g.,][]{ wolfire10, bolatto13, remy15, balashev17},
and abundance ratios, e.g., due to enhanced cosmic ray fluxes
\citep[][]{deforets92, papadopoulos04a, bisbas15, bisbas17}, X-ray
heating from AGN \citep[][]{meijerink07}, or molecule destruction in
shocks \citep[e.g.][]{krips16}. Furthermore, \citet{papadopoulos04a}
and \citet{papadopoulos04b} established the [CI] 1--0 line as a tracer
of gas in high-redshift galaxies that seems to be well mixed with the
molecular gas.

A number of studies have therefore targeted atomic carbon in
high-redshift galaxies, ranging from the ground-work laid by, e.g.,
\citet{brown92}, \citet{barvainis97} and \citet{weiss05} to more
recent studies, in particular of strongly gravitationally lensed
starburst galaxies selected in the infrared \citep[][]{walter11,
alaghband13, bothwell17, yang17, andreani18}. Such studies found luminous line
emission in either one or both lines. They also showed that atomic
carbon can remain optically thin out to the highest gas-mass surface
densities and star-formation rates, and are consistent with high
carbon abundances of a few $\times 10^{-5}$, similar to those found in
low-redshift galaxies
\citep[][]{gerin00,weiss01,israel02,israel03}. This suggests they have
high metallicities akin to the solar value, with little difference
between starburst galaxies and quasars \citep[][]{alaghband13}.  [CI]
combined with other far-infrared and millimeter emission lines also
provides interesting constraints on the gas density and strength of UV
radiation fields within star-forming regions
\citep[e.g.,][]{kaufman99,lepetit06}, and can serve as a discriminant
between the ''starburst'' and ''disk'' modes of star formation, where
the latter is characterized by a larger fraction of diffuse gas
\citep[][]{geach12}.

\begin{figure*}
\centering
\includegraphics[width=0.23\textwidth]{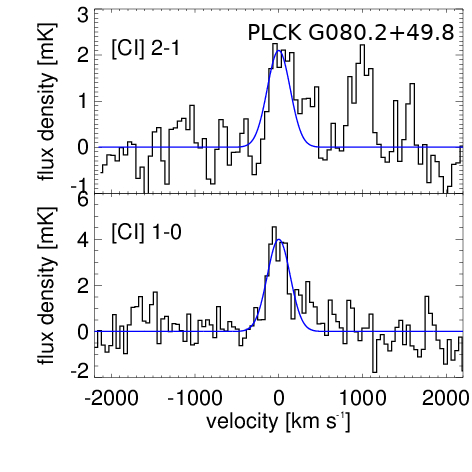}
\includegraphics[width=0.23\textwidth]{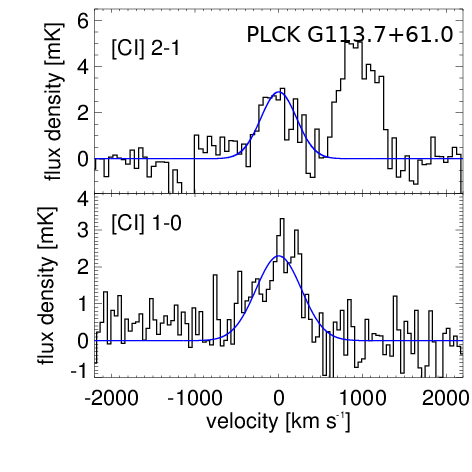}
\includegraphics[width=0.23\textwidth]{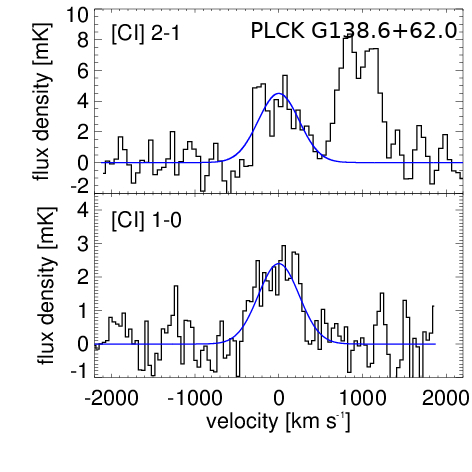}
\includegraphics[width=0.23\textwidth]{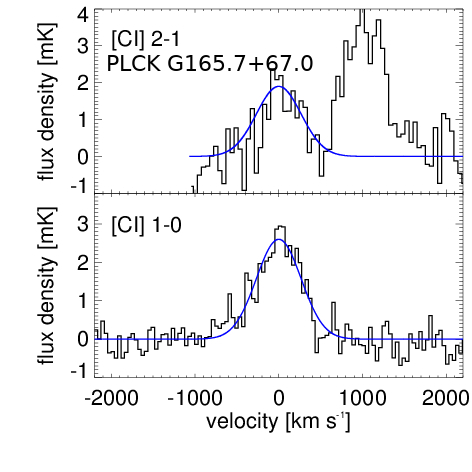}
\caption{[CI] 2--1 and CO(7--6) {\it (top)} and [CI] 1--0 {\it (bottom)}
spectra of the four GEMS for which we observed both lines. The blue
curve shows the single-component Gaussian fit to the [CI] lines. The
upper panel shows also CO(7-6), which is redshifted relative
to [CI] 2--1, and is discussed in detail in \citetalias{canameras18a}.}
\label{fig:ci1021spec}
\end{figure*}
Here we present new observations of [CI] 1--0 and [CI] 2--1 in a small
set of 11 of the brightest gravitationally lensed sub-millimeter
galaxies on the sky observed with the {\it Planck} all-sky survey
\citep[][]{planck15,planck16,dole15}, {\it Planck}'s Dusty GEMS. This sample
is smaller than those found with other infrared-to-millimeter surveys
of high-redshift galaxies \citep[][]{vieira13,wardlow13}, which are
excellent data sets in their own rights, however, the GEMS are particularly
bright dust continuum emitters, reflecting the high completeness
limit of {\it Planck} at $\sim600$~mJy.

All GEMS have spectroscopically confirmed redshifts of $z=2.2-3.6$,
and apparent far-infrared luminosities between 5 and $27\times 10^{13}\ \mu$
L$_{\odot}$ \citep[][]{canameras15}, mainly powered by star formation
and boosted by gravitational lensing from foreground clusters or
massive individual galaxies by luminosity-averaged factors, $\mu \sim
10-30$ \citep[][Canameras et al. 2018b, A\&A accepted, C18 
hereafter]{canameras17b,canameras18a}. Environments probed by these
galaxies range from intense, maximally star-forming clumps
\citep[][]{canameras17a,canameras18a} to diffuse gas probed by [CII]
absorption \citep[][]{nesvadba16}, as observed with ALMA and IRAM
interferometry. AGN contamination is very weak, contributing $\le
10$\% \citep[][]{canameras15} to the infrared luminosity.

The data we present here were obtained as part of several observing
runs with EMIR at the 30-m telescope of IRAM, with the goal of
constraining the global gas properties of these galaxies using several
CO transitions and the two [CI] lines. In total, we observed 48 CO
lines and 15 [CI] lines in 11 galaxies. The results of the analysis of
the CO lines, and of detailed radiative transfer and PDR modeling of
the CO and [CI] lines, are presented in a companion paper \citepalias[][]{canameras18a}. Here
we mainly focus on the physical and empirical properties that can be
derived analytically and from [CI] alone, or that use [CI] for
empirical constraints.

This paper is organized as follows. In Sect.~\ref{sec:obs} we present
the details of our [CI] observations and data reduction. In
Sect.~\ref{sec:lineratios} we describe our analysis and how we
corrected for gravitational lensing, including the possibility of
differential lensing between gas and dust, which we constrain
explicitly using sub-arcsecond interferometry. In
Sect.~\ref{sec:diagnostics} we use the [CI] lines to constrain the
contribution of atomic carbon to line cooling, in order to investigate
the heating mechanism, [CI] excitation temperatures, optical depths,
abundances and masses of atomic carbon. We also determine the gas
distribution and starburst mode from the [CI] line fluxes and line
ratios. In Sect.~\ref{sec:gasmasses} we discuss the implications of
our analysis for H$_2$ gas mass estimates and the CO-to-H$_2$
conversion factor. We summarize our results in
Sect.~\ref{sec:summary}.

Throughout the paper, we adopt the flat $\Lambda$CDM cosmology from
\citet{planck14xvi}, with $H_0$ = 68~km~s$^{-1}$ Mpc$^{-1}$,
$\Omega_{\rm m} = 0.31$, and $\Omega_{\rm \Lambda} = 1 - \Omega_{\rm
  m}$. For example, at $z=3.0$ this implies a luminosity distance 
of 26.0~Gpc, and a projected physical scale of 7.9~kpc~arcsec$^{-1}$.

\section{Observations and data reduction}
\label{sec:obs}

\begin{table*}
\caption{[CI] line properties. We list the source name, the name of
the line, observed frequency, redshift, full-width at half maximum of
the line, main-beam brightness temperature, integrated line flux,
and the line luminosity in units proportional to brightness
temperature and energy, respectively. We give observed values here,
where $\mu$ indicates the gravitational magnification
factor. Luminosity-weighted average gravitational correction factors
are given in Table~1 of \citetalias{canameras18a}, and are repeated in
our Table~\ref{tab:intrinsic} for convenience.\label{tab:lineproperties}}
\begin{tabular}{lcccccccc}
\hline
\hline
Source        & Line    &  $\nu_{\rm obs}$            & redshift           & FWHM         & $\mu\ T_{\rm K}$ &  $\mu\ I_{\rm [CI]}$& $\mu\ L^\prime$ & $\mu\ L$ \\
              &         & [GHz]                  &                    & [km s$^{-1}$] & [mK]         & [Jy km s$^{-1}$]    & [$10^{11}$ K km s$^{-1}$ pc$^2$] & [$10^{8}\ L_{\odot}$] \\
\hline
PLCK$\_$G045.1+61.1   & [CI] 1--0 & 111.133$\pm$0.024      & 3.4280$\pm$0.0002  & 589$\pm$145  & 2.3          & 8.4$\pm$1.7        & 2.3$\pm$0.5    & 8.6$\pm$1.7    \\
\hline
PLCK$\_$G080.2+49.8   & [CI] 1--0 & 136.767$\pm$0.008      & 2.5985$\pm$0.0002  & 242$\pm$61   & 3.8          & 6.2$\pm$0.9        & 1.1$\pm$0.2     & 4.0$\pm$0.7    \\
PLCK$\_$G080.2+49.8   & [CI] 2--1 & 224.847$\pm$0.025      & 2.5995$\pm$0.0003  & 312$\pm$24   & 2.3          & 5.8$\pm$1.0        & 0.37$\pm$0.08    & 6.2$\pm$1.2    \\
\hline
PLCK$\_$G092.5+42.9   & [CI] 1--0 & 115.641$\pm$0.016      & 3.2559$\pm$0.0006  & 475$\pm$128  & 4.5          & 13.5$\pm$2.6       & 3.3$\pm$0.6      & 12.8$\pm$2     \\
\hline
PLCK$\_$G102.1+53.6   & [CI] 2--1 & 206.608$\pm$0.006     & 2.9173$\pm$0.0001  & 220$\pm$21   & 2.3          & 4.0$\pm$0.4        & 0.3$\pm$0.3    & 5.2$\pm$0.5   \\
\hline
PLCK$\_$G113.7+61.0   & [CI] 1--0 & 144.019$\pm$0.020      & 2.41730$\pm$0.0003 & 639$\pm$100  & 2.3          & 10.2$\pm$1.2       & 1.5$\pm$0.2    & 5.9$\pm$0.7   \\
PLCK$\_$G113.7+61.0   & [CI] 2--1 & 237.395$\pm$0.002      & 2.40927$\pm$0.0001 & 504$\pm$10   & 2.9          & 9.2$\pm$ 1.0       & 0.5$\pm$0.05   & 8.6$\pm$0.9   \\
\hline 
PLCK$\_$G138.6+62.0   & [CI] 1--0 & 142.974$\pm$0.020      & 2.4423$\pm$0.0003  & 575$\pm$86   & 2.4          & 9.5 $\pm$1.5       & 1.5$\pm$0.2    & 5.6$\pm$0.9    \\
PLCK$\_$G138.6+62.0   & [CI] 2--1 & 235.129$\pm$0.003      & 2.4421$\pm$0.0001  & 526$\pm$5    & 4.5          & 18.8$\pm$0.2       & 1.1$\pm$0.01    & 18.0$\pm$0.1   \\
\hline
PLCK$\_$G145.2+50.9   & [CI] 1--0 & 108.204$\pm$0.009      & 3.5485$\pm$0.0003  & 405$\pm$63   & 5.8          & 14.8$\pm$3.5       & 4.2$\pm$1.0    & 16.0$\pm$3.8    \\
\hline
PLCK$\_$G165.7+67.0   & [CI] 1--0 & 152.079$\pm$0.009      & 2.2362$\pm$0.0001  & 629$\pm$46   & 2.6          & 11.1$\pm$2.8       & 1.5$\pm$0.4    & 5.6$\pm$1.4     \\
PLCK$\_$G165.7+67.0   & [CI] 2--1 & 250.059$\pm$0.025      & 2.2366$\pm$0.0002  & 418$\pm$6    & 2.6          & 8.3 $\pm$0.2       & 0.4$\pm$0.09   & 6.8$\pm$0.1    \\
\hline 
PLCK$\_$G200.6+46.1   & [CI] 2--1 & 203.697$\pm$0.018      & 2.97326$\pm$0.0001 & 412$\pm$5    & 2.5          & 8.3$\pm$0.1        & 0.7$\pm$0.01   & 11.1$\pm$0.1    \\
\hline
PLCK$\_$G231.3+72.2   & [CI] 2--1 & 209.729$\pm$0.015      & 2.85899$\pm$0.0001 & 319$\pm$18   & 2.7          & 6.9$\pm$0.4        & 0.5$\pm$0.03   & 8.7$\pm$0.5 \\
\hline
PLCK$\_$G244.8+54.9   & [CI] 2--1 & 202.113$\pm$0.002      & 3.00440$\pm$0.0001 & 586$\pm$5    & 4.5          & 21.0$\pm$0.2       & 1.7$\pm$0.02   & 28.6$\pm$0.3 \\
\hline
\hline
\end{tabular}
\end{table*}

\subsection{IRAM/EMIR spectroscopy}
\label{ssec:spectroscopy}

We obtained deep spectroscopy of several bright millimeter emission
lines, including the [CI] lines presented here, with the wide-band
millimeter receiver EMIR at the 30-m telescope of IRAM in several runs
between November~2012 and February~2015. In total, we
obtained between 81~min and 171~min of on-source observing time per
source. Individual observing dates and integration times, tuning
frequencies, and rms noise values are given in Table~\ref{tab:observations}
for each source and emission line. The analysis of the CO lines
is presented in \citetalias{canameras18a}.

Depending on the redshift of each source, the
[CI] 1--0 line either falls into the 3--mm or 2--mm band, and the [CI] 2--1
line either into the 2--mm or the 1.3--mm band. In two sources,
PLCK$\_$G113.7$+$61.0 and PLCK~G138.6$+$62.0, the [CI] 1--0 line was used
to confirm the spectroscopic redshift previously estimated from a
blind line search in the 3--mm band \citep[][]{canameras15}. In most
cases, the [CI] lines were observed with dual-band observations,
i.e. in parallel to other bright millimeter emission lines.

 Data were taken under a range of atmospheric conditions. For the 3--mm
observations, preciptable water vapor columns (pwv) were mostly
between 1~mm and 8~mm. A small part of the observing time suffered
from even higher pwv; including these scans did not improve the
signal-to-noise ratios in the final combined data sets, and so these
scans were discarded. Observations at 1.3~mm were carried out when the
pwv was 1~mm or less.

We used the FTS and WILMA backends with Wobbler switching throws of
60\arcsec, which is significantly larger than the diameter of our most
extended sources, about 10\arcsec. To point the telescope we used
blind offsets from radio-loud quasars at distances of a few degrees
from our targets. We performed a pointing approximately once every
2~hr, and refocused the telescope every 3-4 hr, and at sunrise and
sunset.  Individual scans were 30~seconds long, and we obtained a
calibration after every 6~minutes of observing. The FTS and WILMA
backends have intrinsic resolutions of 0.195 and 2~MHz, respectively,
and 16~GHz and 8~GHz of bandwidth, respectively, with horizontal and
vertical polarizations observed in parallel. We typically rebinned the
data to more appropriate spectral resolutions between 30 and 50 km s$^{-1}$
(see Figs.~\ref{fig:ci10spec} and ~\ref{fig:ci21spec}).

All data were calibrated at the telescope and reduced with the CLASS
package of the GILDAS software of IRAM \citep[][]{gildas13}. We
inspected all individual scans by eye and used simple first-order
polynomials to correct the baselines, after carefully masking the
spectral range expected to be covered by the bright emission
lines. For scans with strong 'platforming' in their FTS spectra, we
used the routine {\tt FtsPlatformingCorrection5.class} kindly provided
by C.~Kramer to obtain individual spectral scans with reasonably flat
baselines. We used the values given on the EMIR
website\footnote{http://www.iram.es/IRAMES/mainWiki/Iram30mEfficiencies}
to approximate the antennae efficiency and to translate the measured
brightness temperatures into flux density units (Jy).

\section{Line measurements}
\label{sec:lineratios}

We detected all targeted [CI] 1--0 and [CI] 2--1 lines, i.e., all [CI]
lines from galaxies in this sample that fall into the atmospheric
windows, with line fluxes between 3.4 and 21 Jy km s$^{-1}$ and FWHM
line widths between 220 and 640~km s$^{-1}$.
We used the {\sc CLASS} function {\sc line} for an initial line fit
after subtracting the baseline \citep[][]{gildas13}, which we then confirmed with the {\sc
mpfit} routine using IDL~\citep[][]{markwardt09}. Within the
limits imposed by different signal-to-noise ratios, our fit results
are consistent with those obtained for CO by \citetalias{canameras18a}.

We followed, e.g., \citet{solomon05} to calculate emission-line
luminosities, $L_{\rm line}$ and $L^\prime_{\rm line}$, from these fluxes by setting
\begin{equation}
L_{\rm line}=1.04\times 10^{-3}\ S_{\rm line}\ \Delta v\ \nu_{\rm rest}\ (1+z)^{-1}\ D^2_{\rm L}
\end{equation}
where $S_{\rm line} \Delta v$ is the velocity-integrated line flux given
in Jy km s$^{-1}$, and $\nu_{rest}$ the frequency in the rest-frame in
GHz. $z$ is the redshift. $D_{\rm L}$ is the luminosity distance in Mpc. $L_{\rm line}$
is given in solar luminosities. An alternative way to express line
luminosities is by setting

\begin{equation}
L^\prime_{\rm line} = 3.26\times 10^7\ S_{\rm line}\Delta v\ \nu_{\rm obs}^{-2}\ D_L^2\ (1+z)^{-3}
\end{equation}
where $\nu_{\rm obs}$ is the observed frequency. The resulting
luminosities, $L^\prime$, are proportional to the brightness
temperature and are used, for example, to calculate gas
masses. $L^\prime$ is given in K km s$^{-1}$ pc$^2$.

\begin{figure*}
\centering
\includegraphics[width=0.3\textwidth]{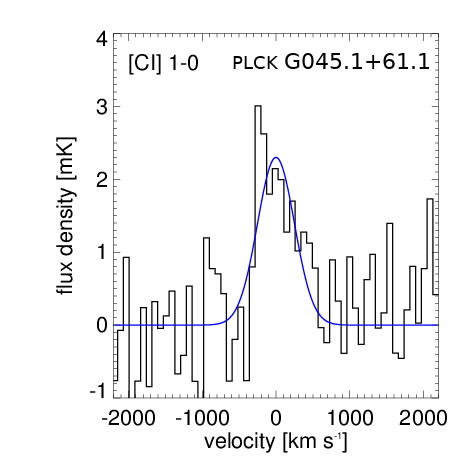}
\includegraphics[width=0.3\textwidth]{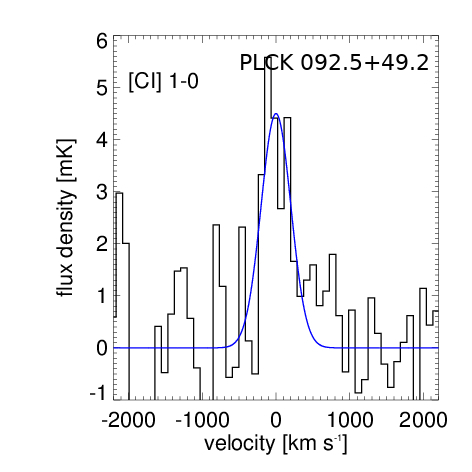}
\includegraphics[width=0.3\textwidth]{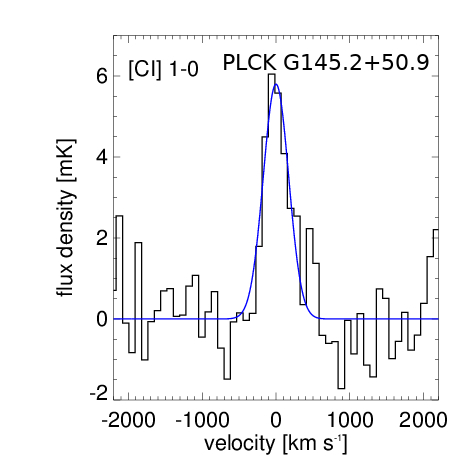}
\caption{[CI] 1--0 line of the GEMS for which
[CI] 2--1 and CO(7-6) fall outside the atmospheric windows. Blue curves indicate the single-component Gaussian fits to the [CI] 1--0 line.}
\label{fig:ci10spec}
\end{figure*}

\begin{figure*}
\centering
\includegraphics[width=0.3\textwidth]{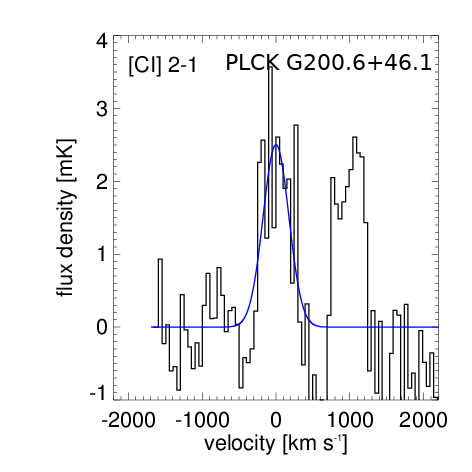}
\includegraphics[width=0.3\textwidth]{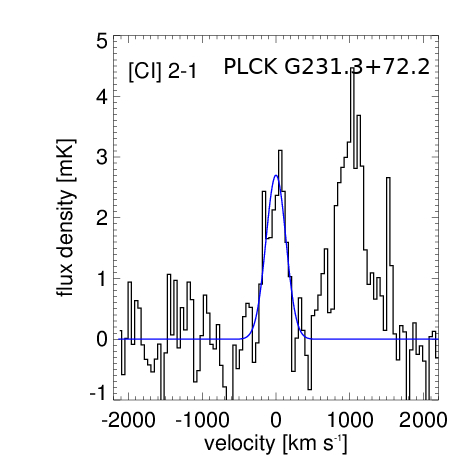}
\includegraphics[width=0.3\textwidth]{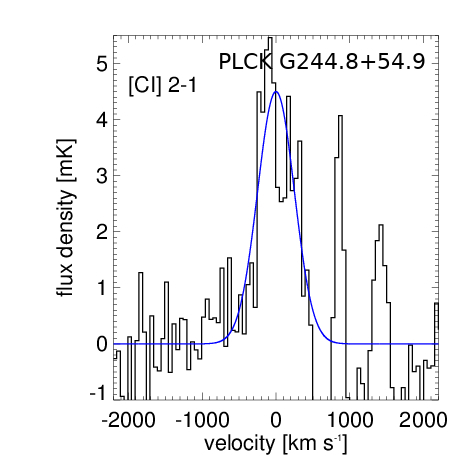}
\caption{[CI] 2--1 and CO(7-6) lines of the GEMS
for which [CI] 1--0 falls outside the atmospheric
windows. For PLCK$\_$G244.8$+$54.9, [CI] 2--1 falls right at the edge of the band, so
that [CI] 2--1 can be reliably measured, but CO(7-6) cannot. Blue curves
indicate the single-component Gaussian fits to the [CI] 2--1 line.}
\label{fig:ci21spec}
\end{figure*}

In total, we measured [CI] 1--0 in seven galaxies, and [CI] 2--1 in
eight. This includes four galaxies where we measured both [CI] lines
(Fig.~\ref{fig:ci1021spec}). Individual line profiles are shown in
Fig.~\ref{fig:ci10spec} for [CI] 1--0 and in Fig.~\ref{fig:ci21spec}
for [CI] 2--1. CO(7--6) has a rest frequency of 806.6518~GHz, only
2.7~GHz (or about 1000 km s$^{-1}$) to the red from the [CI] 2--1 line
at 809.3435~GHz. The figures showing the [CI] 2--1 lines therefore
also cover the CO(7--6) line in all cases except one. In
PLCK$\_$G244.8$+$54.9, [CI] 2--1 falls right at the band edge; it can
be measured with a reliable calibration, unlike CO(7--6). For a
detailed discussion of the line properties of CO(7--6) and other CO
lines, see \citetalias{canameras18a}.

The line profiles are generally smooth enough to be well fitted with
single Gaussian profiles (Figs.~\ref{fig:ci1021spec} to
\ref{fig:ci21spec}). In three of the four galaxies where both
[CI] 1--0 and [CI] 2--1 were measured, the profiles of both lines are
similar within the signal-to-noise ratio of the present data, and the
line centers are at similar redshifts for both transitions (see
Fig.~\ref{fig:ci1021spec}). In the fourth, PLCK\_G113.7$+$61.0,
[CI]1--0 and [CI]2--1 are offset by 700 km s$^{-1}$, comparable to the
FWHM of the [CI] 1--0 line (639$\pm$100,
Table~\ref{tab:lineproperties}), and perhaps indicating velocity
structure within the galaxy. In PLCK\_G045.1$+$61.1,
PLCK\_G092.5$+$42.9, and PLCK\_G244.8$+$54.9, the signal-to-noise
ratios are not high enough to fit two separate line components like
done for the brightest CO lines (C18). The results of our line fits are listed
in Table~\ref{tab:lineproperties}, and they are not corrected for
gravitational lensing. Luminosity-weighted magnification factors are
given in Table~1 of \citetalias{canameras18a}.

\subsection{Gravitational magnification and differential lensing}
\label{ssec:difflensing}

We have constructed detailed gravitational lens models for all GEMS
\citep[][Canameras et al. 2018c, in prep.]{canameras17b,canameras18a},
which we derived with the publicly available {\sc Lenstool} package
\citep[][]{jullo09}. {\sc Lenstool} models the lensing
potentials as pseudo-isothermal ellipsoids and derives
the properties of these ellipsoids by calculating the expected
position of multiple gravitationally lensed arclets behind the lensing
structure. We used the HST/WFC3 imaging recently presented by
\citet{frye18}, for the five sources where it was available, and
ground-based CFHT imaging with 0.8\arcsec\--1.0\arcsec\ resolution
otherwise. Residuals between observed and modeled positions of arclets
are smaller than the size of the PSF in all models.

From the detailed lensing models, which we constrained from the WFC3
morphologies, and sub-arcsecond millimeter dust and CO emission-line
maps and the kinematic properties of the gas in each source (and which
therefore take into account the source morphology and differential
lensing) we calculate luminosity-weighted average magnification
factors separately for the gas and the dust, finding values between 6
and 30 \citepalias[see Table~1 of][]{canameras18a}. Deriving
average magnifications for the dust and gas from pixel-by-pixel
reconstructions of the source-plane image suggests uncertainties from
differential lensing of about 25\%.

PLCK$\_$G138.6$+$62.0 is the only galaxy for which we do not have
spatially resolved millimeter or sub-millimeter morphologies, so we
adopt an empirically estimated factor $\mu=20$ in this case \citep[for
details see ][]{canameras15}. For the [CI] line we are most concerned
with here, we use the magnification factors derived from the CO line
emission, i.e., assuming that both CO and [CI] come from gas clouds
with similar spatial distributions. We thus neglect a potential
contribution from faint, very extended diffuse gas outside the bright
star-forming regions themselves. This assumption can be tested
indirectly by comparing the line profiles, which indeed do not show
significant differences when integrated over entire sources. The
values we used for this paper are listed in Table~\ref{tab:intrinsic}.

We can also use the different estimates for the dust and gas masses to
roughly constrain the impact of differential lensing on the various
lines in the sub-millimeter and millimeter regime, finding rather moderate
typical differences of about 25\%, without any dramatic outliers. This
is also to be expected, given that the dust and the mid-J CO and [CI]
2--1 lines should mainly originate from gas and dust in actively
star-forming regions \citep[and noting that the GEMS do not show
evidence of luminous AGN]{canameras15}.

\begin{figure*}
  \includegraphics[width=0.48\textwidth]{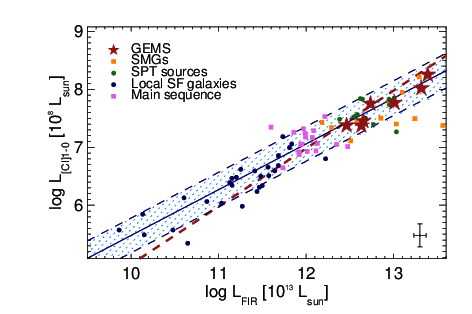}
  \includegraphics[width=0.48\textwidth]{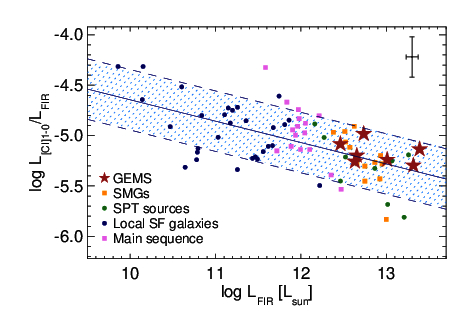}
  \caption{\label{fig:cicooling} Cooling budget through [CI]1--0 line
emission. [CI]1--0 luminosity {\it (left panel)}, $L_{CI1-0}$, and
ratio with far-infrared luminosity, L$_{CI1-0}$/L$_{\rm FIR}$ {\it
(right panel)}, as a function of far-infrared luminosity, L$_{\rm FIR}$.
The red stars are the GEMS. Blue, pink, green, and yellow symbols
indicate the samples of low-redshift spiral galaxies from
\citet{kamenetzky16}, the main sequence galaxies at $z\sim1.2$ from
\citet{valentino18}, the high-redshift samples of gravitationally
lensed sources from the South Pole Telescope \citep{bothwell17} and
the submillimeter galaxies of \citet{alaghband13}, respectively. The
blue line shows the average relationships derived by
\citet{valentino18} for their main-sequence galaxies and comparison
samples, and the red line shows an equivalent relationship with a
slope of unity. Blue hatched bands show a range of~$\pm 2$ around
these averages. Typical error bars of our measurements are shown in
the lower and upper right corners of the two panels, respectively.}
\end{figure*}

We can also constrain the likely impact of differential lensing
directly from the present data. As we will show in
Sect.~\ref{ssec:atomiccooling}, the cooling budget from [CI] relative
to the far-infrared luminosity and to CO(4--3) is within a factor of~2
from that found in other samples of nearby and high-redshift galaxies,
including gravitationally lensed and unlensed galaxies
(Fig.~\ref{fig:cicoratio}, and Fig.~\ref{fig:cicooling}),
respectively. In the absence of a systematic conspiracy with other
astrophysical quantities, this suggest that differential lensing does
not introduce larger uncertainties than other effects. Moreover,
integrated mass estimates from CO(1--0) and [CI] 1--0 are very
similar, as we will show in Sect.~\ref{sec:gasmasses}.

For the same reason, we consider it unlikely that we have missed a
dominant component of CO-dark, [CI] 1--0 -emitting gas that is
strongly gravitationally lensed and has significant transversal
positional or velocity offsets from the molecular clouds. However,
this does not imply that CO and [CI] are exactly co-spatial
\citep[e.g.,][]{offner14}. On scales of a few hundred parsecs or less,
smaller than the area that is being magnified by the gravitational
lens, and in directions roughly along the line of sight, or
perpendicular to the magnification direction, the diffuse and dense
gas may or may not be well mixed, without impact on differential
lensing. This would be the case, e.g., for a clumpy interstellar
medium. Sizes of-order 100~pc are consistent, e.g., with the
Jeans-length in dense, fragmenting gas disks of high-z galaxies
including the GEMS \citep[][C18]{hodge18, canameras17a, swinbank11}.

\section{[CI] diagnostic properties} 
\label{sec:diagnostics}

With the line fluxes measured in Sect.~\ref{sec:lineratios}, we can
derive luminosity ratios between the [CI] 2--1 and [CI] 1--0 line for
the four galaxies, in the cases where both lines are measured. We find
ratios of $L_{{\rm CI2-1/CI1-0}}= 1.2$ to $3.3$
(Table~\ref{tab:lineproperties}). Given the wide range in gas
conditions probed by the [CI] lines, their ratios with each other,
with other emission lines (in particular CO), and with the
far-infrared continuum, all provide interesting diagnostic
constraints.

A thorough analysis of the gas excitation using the radiative transfer
code {\sc RADEX} and PDR model of \citet{kaufman99} has already been
presented by \citetalias{canameras18a}, using [CI] as well as multiple CO lines, generally
between $J=3 - 2$ and $J=7 - 6$, and evan above J=$9 - 8$ for two galaxies.
They find that the gas in the GEMS is characterized by
luminosity-weighted average gas densities of $n\sim
10^{4-5}$~cm$^{-3}$, and radiation fields of $10^{2-4}\ G_0$; these are
in the range of other starburst galaxies at low and
high redshift. Here we will complement and extend these analyses by
focusing on the constraints that can be derived solely from the atomic
carbon lines, as well as several empirical constraints on the gas
masses and the distribution of interstellar gas in the GEMS.

\subsection{Atomic line cooling}
\label{ssec:atomiccooling}
We can use the [CI] line luminosities, $L_{\rm CI}$, and the far-infrared
luminosities from \citet{canameras15} integrated over a wavelength
range of $8-1000$ $\mu$m, to estimate the total cooling from atomic
gas in the GEMS. Using the luminosities $L_{\rm [CI1-0}$ and $L_{\rm [CI]
2-1}$ listed in Table~\ref{tab:lineproperties}, we find
$L_{\rm CI1-0}/L_{\rm FIR}$ and $L_{\rm CI2-1}/L_{\rm FIR}$ ratios of 5 to
$20\times 10^{-6}$. We adopted the measured values and did not correct
for differences in dust and gas magnifications, which would have
changed our results by at most about 25\%. Values for individual
galaxies are given in Table~\ref{tab:masses}.

\citet{bothwell17} found $L_{\rm CI1-0}/ L_{\rm
FIR}=7.7\pm2.4\times 10^{-6}$ in their sample of 13 gravitationally
lensed galaxies at $z\sim 4$ from the SPT survey. \citet{walter11}
measured $L_{\rm CI1-0}/L_{\rm FIR}= 1-15\times 10^{-6}$, albeit
using far-infrared luminosities that were derived prior to the launch
of the {\it Herschel} satellite, from the flux density at 850~$\mu$m
and a fiducial dust temperature of $T=35$~K. \citet{alaghband13} found
$L_{\rm CI1-0}/L_{\rm FIR}=2.6\pm 0.5\times 10^{-5}$ for their
newly observed sources, and $L_{\rm CI1-0}/L_{\rm FIR}=8 \times
10^{-6}$ for sources culled from the literature \citep[which have
considerable overlap with the sample of ][]{walter11}. 
\citet{valentino18} find about half the value, $L_{\rm [CI]
1-0}/L_{\rm FIR}=1.4\times10^{-5}$ for a set of main-sequence galaxies
at z=1.2. The GEMS therefore fall within the wide range previously
found in other high-redshift galaxies. This can also be seen from
Fig.~\ref{fig:cicooling}, where we plot $L_{\rm CI}$ and
$L_{\rm CI}/L_{\rm FIR}$ as a function of $L_{\rm FIR}$. The GEMS follow
similar trends as the samples of low and high-redshift galaxies. All
fall within a factor of~2 of the best-fit relations.

In nearby ULIRGs, \citet{rosenberg15} find that the combined ratio of
the two [CI] lines is $L_{CI10+21}/L_{\rm FIR}=1-5\times 10^{-5}$ in
most galaxies, except for the lowest FIR luminosities in the LIRG
regime, where ratios can reach about $1\times 10^{-4}$. The GEMS with
both lines measured have combined ratios $L_{CI10+21}/L_{\rm FIR}=
1.2$ and $2.6\times 10^{-5}$, in the lower range found in the nearby
Universe; we find similar values when using the [CI] 2--1 / [CI] 1--0
ratios of the GEMS as a fiducial correction factor of the missing
second line in the samples of \citet{bothwell17} and
\citet{alaghband13}. \citet{riechers13} found L$_{{\rm
CI10+21}}/L_{\rm FIR}=2-5\times 10^{-5}$ in a luminous starburst at
z=6.3.

This suggests, at least for the small sample sizes and signal-to-noise
ratios obtained for current samples of high-redshift galaxies, that
the contribution of atomic gas to the overall cooling budget of the
galaxies has remained approximately constant since about 1~Gyr after
the Big Bang, and at most slightly increased with cosmic time. In nearby
galaxies, the two [CI] lines contribute together about 1.5\% to the
total gas cooling rate \citep[][]{rosenberg15}.

Finding similar line-to-continuum flux ratios in high and low-redshift
galaxies imposes at least loose upper limits on the importance of
global changes in the gas heating processes in high-redshift galaxies
due to cosmic rays \citep[][]{bisbas17} or X-rays
\citep[][]{meijerink07}. If, on top of the heating from UV photons,
bolometrically significant, additional heating mechanisms like X-rays,
cosmic rays, or shocks were present (which predominantly boost the
line, but not the continuum luminosities at long wavelengths), we
would expect these ratios to be systematically greater in
high-redshift galaxies. Our results suggest that this is not the
case. Given the scatter in the relationships, however, this does not
imply that such mechanisms are not present, they just cannot
dominate the overall gas heating budgets.

\subsection{Heating mechanism and AGN contamination}
\label{ssec:agnxdr}

The ratio of the [CI] 2--1 and the [CI] 1--0 line fluxes can inform us
about the presence of X-ray heating from AGN. Radiative transfer
models of gas heating from UV photons imply an upper limit to the [CI]
2--1 / [CI] 1--0 ratio, which cannot be exceeded without the presence
of a harder incident radiation field like that from an
AGN. \citet{meijerink07} calculated line ratios for gas heated by UV
and X-ray photons, as expected for regions of intense star formation
and circum-nuclear environments within AGN host galaxies,
respectively. For a wide range in gas density between about 10 and
$10^6$ cm$^{-3}$, they predict that X-ray heating will produce line
ratios between [CI] 2--1 and [CI] 1--0 of $L_{21}/L_{10}\ga
3.5$. Ratios lower than this are a clear indication of UV heating in
gas with typical densities of a few $\times 10^{2-4}$ cm$^{-3}$ as in
the GEMS \citepalias[][]{canameras18a}.

In Table~\ref{tab:text} we give the luminosity ratios for the four
galaxies where we measured both [CI] lines. In PLCK$\_$G080.2$+$49.8,
PLCK$\_$G113.7$+$61.0, and PLCK$\_$G165.7$+$67.0, we find very similar
ratios, between $L_{[\rm CI21}/L_{\rm CI10}=1.2\pm0.3$ and
$L_{\rm CI21}/L_{\rm CI10}=1.5\pm0.2$. However, the ratio in the fourth source,
PLCK$\_$G138.6$+$62.0, is significantly higher,
$L_{\rm CI21}/L_{\rm CI10}=3.3\pm0.1$. This source therefore falls near the regime
where an AGN could have an impact, although it is still within the
range expected for intensely star-forming systems. Finding little
evidence for AGN X-ray heating from the line ratios confirms our
previous results from the mid-to-far-infrared spectral energy
distributions, which also suggest that AGN are weak compared to the UV
radiation from young stellar populations \citep[][]{canameras15}, or
absent.

\subsection{Star-formation mode}
\label{ssec:sfmode}

\citet{greve12} and \citet{papadopoulos12} proposed that the ratio of
the line luminosities of CO(4--3) and [CI] 1--0 can be used to infer
qualitatively the relative amount of dense molecular and more diffuse
gas. They associate higher ratios of dense molecular to diffuse gas
with starburst galaxies, and galaxies with a more balanced ratio of
dense and diffuse gas with quiescently star-forming (disk-like)
galaxies. From observations of nearby galaxies, they infer an average
ratio of $r_{\rm CO(4-3)/CI10}=4.55\pm 1.5$ for starburst (ULIRG)
environments, and $r_{\rm CO(4-3)/CI10}=0.45-1.3$ for disk
galaxies. In the GEMS, the corresponding ratios are between 2.6 and
5.8.  For galaxies where we have a direct measurement of [CI] 1--0, we
find ratios of 2.9$-$3.3, and 2.6$-$5.8 for the galaxies without [CI]
1--0 measurement (where we used the [CI] 2--1 measurement corrected for
an average [CI] 2--1/[CI] 1--0 ratio instead). These results are all in
the starburst regime, as also expected from the high star-formation
rate densities found by \citet{canameras17a}.

We note that using [CI] 2--1 instead of [CI] 1--0 can lead to
uncertainities of factors of 2-3, and additionally for some galaxies
we used CO(3--2) instead of CO(4--3), because CO(4--3) falls outside the
atmospheric window (see Table~\ref{tab:ratios}). From the CO spectral
line energy diagrams shown by \citetalias{canameras18a}, we know that
this might bias the line ratios of the GEMS about 25\% low compared to
estimates with CO(4-3). Since we only aim at loosely classifying the
GEMS between two groups that differ by an order of magnitude on
average, and do not use the precise value of these line ratios, we
find that our conclusions are not compromised by these additional
systematic uncertainties.
\begin{table}
\begin{center}
\begin{tabular}{lcccc}
\hline \hline
Source        & ${\rm M}_{\rm CI}$ & ${\rm M}_{\rm H_2}$  & $\mu_{\rm gas}$ \\
              & [$10^7$ M$_{\odot}$] & [$10^{10}$ M$_{\odot}$] & \\
\hline
PLCK$\_$G045.1$+$61.1 & 1.9     & 13.5      & 15.5 \\
PLCK$\_$G080.2$+$49.8 & 0.8     & 6.2       & 15.9 \\
PLCK$\_$G092.5$+$42.9 & 3.7     & 25.8      & 12.0 \\
PLCK$\_$G113.7$+$61.0 & 2.0     & 14.6      & 9.7  \\
PLCK$\_$G138.6$+$62.0 & 1.0     & 6.8       & 20.  \\
PLCK$\_$G145.2$+$50.9 & 6.3     & 43.7      & 8.9  \\
PLCK$\_$G165.7$+$67.0 & 0.8     & 5.8       & 24.1 \\
\hline \hline
\end{tabular}
\end{center}
\caption{\label{tab:intrinsic} Intrinsic masses of atomic carbon,
${\rm M}_{\rm CI}$, and molecular gas, ${\rm M}_{\rm H_2}$, estimated from [CI]
1--0 in the seven galaxies where we observed this line. For
convenience, we also list the magnification factors for the gas,
$\mu_{\rm gas}$, taken from Table~1 of \citetalias{canameras18a}, which we have used to
correct these measurements for gravitational lensing.}
\end{table}

\subsection{Excitation temperatures and optical depth}
\label{ssec:optdepth}

One of the main advantages in using [CI] instead of CO lines as a
tracer of mass is that they should remain optically thin out to the
high volume-averaged column densities typically encountered in rapidly
star-forming dusty high-redshift galaxies
\citep[e.g.,][]{walter11}. In other words, the line luminosity remains
proportional to the total mass. Since we have both the [CI] 2-1 and
[CI] 1-0 line measured in four GEMS, we can test this
assumption directly.

We just saw in Sect.~\ref{ssec:agnxdr} that the gas in the present galaxies is
predominantly heated by UV photons and we can therefore follow
\citet{schneider03} and \citet{walter11}, who derived the optical depth, $\tau$, of the
[CI] 1--0 emitting gas in photon-dominated regions (PDRs) by setting
\begin{equation}
\tau_{\rm [CI]1-0} = -\ln{(1-T_{\rm mb,[CI]1-0}\ ({\rm e}^{23.6/T_{\rm ex}}-1)/23.6)}.
\end{equation}
Here $T_{\rm ex}$ is the excitation temperature of the gas in kelvin,
$K$, assuming LTE.  $T_{\rm mb}$ is the rest-frame peak
intensity of the line in main beam brightness temperature, and is also
given in kelvin. A similar expression can be given for [CI] 2--1:
\begin{equation}
\tau_{\rm [CI]2-1} = -\ln{(1-T_{\rm mb,[CI]2-1}\ (e^{38.8/T_{ex}}-1)/38.8)}. 
\end{equation}

\begin{table*}
\caption{\label{tab:text} Ratios of line luminosities to the far-infrared luminosity of [CI] 1--0 and [CI] 2--1, and luminosity ratios of [CI] 2--1 and [CI] 1--0. Excitation temperature and optical depths of [CI] 1--0 and [CI] 2--1 are also given, as well as the mass of atomic carbon (not corrected for gravitational magnification $\mu$) for galaxies with [CI] 1--0 observation.}
\begin{center}
\begin{tabular}{lcccccccc}
\hline \hline
  Source                & $ L_{\rm CI10}$$/L_{\rm FIR}$ & $L_{\rm CI21}/L_{\rm FIR}$ & $L_{\rm CI21}/L_{\rm CI10}$& $L^\prime_{\rm CI21}/L^\prime_{\rm CI10}$   & $T_{\rm ex}$             & $\tau_{CI1-0}$ & $\tau_{CI2-1}$ &  $\mu {\rm M}_{\rm CI}$          \\
                        &  [$\times 10^{-6}$]     &  [$\times 10^{-6}$]     &                        &                                     &  [K]                 &                &                &  [$10^8$ M$_{\odot}$]  \\
  \hline
  PLCK$\_$G045.1$+$61.1 &  10.2$\pm$2.0          & \dots       & \dots                   & \dots                               & 20\tablefootmark{a}  & 0.04\tablefootmark{a} & \dots   & 3.0$\pm$0.5            \\
  PLCK$\_$G080.2$+$49.8 &  8.7$\pm$1.5          & 13.5$\pm$2.6 &  1.6$\pm$0.2            &  0.33$\pm$0.1                       & $22.3^{+3.75}_{-3.5}$   & 0.14           & 0.05           & 1.3$\pm$0.3           \\
  PLCK$\_$G092.5$+$42.9 &  5.2$\pm$0.8           & \dots       &  \dots                  & \dots                               & 20\tablefootmark{a}  & 0.07\tablefootmark{a} & \dots   & 4.4$\pm$0.3           \\
  PLCK$\_$G102.1$+$53.6 &  \dots                 & 6.6$\pm$0.6 & \dots                   & \dots                               & \dots                & \dots & 0.02\tablefootmark{a}   & \dots                 \\
  PLCK$\_$G113.7$+$61.0 &  6.0$\pm$0.7           & 8.6$\pm$0.9 &  1.5$\pm$0.2            & 0.33$\pm$0.2                        & $21.0^{+8.7}_{-7.3}$    & 0.03           & 0.02           & 2.0$\pm$0.4           \\
  PLCK$\_$G138.6$+$62.0 &  6.1$\pm$1.0           & 20.0$\pm$0.1 &  3.3$\pm$0.1            & 0.73$\pm$0.2                        & $36.7^{+10.5}_{-8.3}$   & 0.01           & 0.01           & 1.9$\pm$0.3           \\
  PLCK$\_$G145.2$+$50.9 &  7.3$\pm$1.7           & \dots       &  \dots                  & \dots                               & 20\tablefootmark{a}  & 0.02\tablefootmark{a} &  \dots  & 5.6$\pm$1.0           \\
  PLCK$\_$G165.7$+$67.0 &  5.4$\pm$1.3           & 6.6$\pm$0.1 &  1.2$\pm$0.3            & 0.27$\pm$0.1                        & $18.7^{+3.38}_{-3.5}$   & 0.02           & 0.01           & 2.0$\pm$0.3           \\ 
  PLCK$\_$G200.6$+$46.1 &  \dots                 & 19.5$\pm$0.2 & \dots                   & \dots                               & \dots                &  \dots  & 0.01\tablefootmark{a} & \dots                 \\
  PLCK$\_$G231.3$+$72.2 &  \dots                 & 11.6$\pm$0.7 & \dots                   & \dots                               & \dots                &  \dots         & 0.06\tablefootmark{a}                & \dots \\
  PLCK$\_$G244.8$+$54.9 &  \dots                 & 10.8$\pm$0.1 & \dots                   & \dots                               & \dots                &  \dots         & 0.55\tablefootmark{a} & \dots          \\
 \hline \hline
\end{tabular}
\end{center}
\tablefoottext{a}{For galaxies without either [CI] 1--0 or [CI] 2--1 measurement, we adopted a fiducial excitation temperature of $T_{\rm ex}=$20 K.}
\end{table*}

The excitation temperature in kelvin can be found from the ratio of
line luminosities, $L^\prime_{CI2-1} / L^\prime_{CI1-0}$, by
setting
\begin{equation}
T_{\rm ex} = h \nu_{21} / k_{\rm B}\ \ln (\frac{N_{10}}{N_{21}} \frac{g_{21}}{g_{10}} )^{-1} = \frac{38.8}{\ln(2.11/R)}\ [{\rm K}],
\end{equation}
where $R$ is the ratio between the integrated luminosities,
$L^\prime$, of the [CI] 2--1 and [CI] 1--0 lines, $k_{\rm B}$ is the
Boltzmann constant. $h$ is Planck's constant, $\nu_{21}$ the
rest-frame frequency of the [CI] 2--1 line, $N_{10}$ and $N_{21}$ are
the column densities of the [CI] 1--0 and 2--1 line, respectively, and
$g_{21}$ and $g_{10}$ are the corresponding Gaunt factors.

With the luminosities and main-beam brightness temperatures given in
Table~\ref{tab:lineproperties}, we find excitation temperatures,
$T_{\rm ex}=21-37$~K. This is consistent with previous work
\citep[][]{jiao17} and systematically lower than the dust temperatures
found by \citet{canameras15}, which are between 33 and 50~K for the
same galaxies; this might indicate that [CI] has a significant
extended component (see also Sect.~\ref{sec:gasmasses}), or that the
dust and atomic gas are not in thermal equilibrium.

Both lines are optically thin in the GEMS, and comparable to those in
other high-redshift galaxies \citep[][]{walter11, alaghband13}. The
corresponding optical depths of the [CI] 1--0 line are between
$\tau_{10}=0.01$ and $0.14$, and for the [CI] 2--1 line generally
between $\tau_{21}=0.01$ and $0.06$. In PLCK$\_$G244.8$+$54.9 we find
$\tau_{21}=0.55$ for a fiducial temperature of $T_{\rm ex}=$20~K. This
temperature is likely too low for a galaxy with such highly excited
gas \citepalias{canameras18a}. For $T_{\rm ex}=$40~K, we would find a
more typical value of $\tau_{21}=0.13$. Results for individual
galaxies are listed in Table~\ref{tab:text}; in galaxies where only
one [CI] line falls into the atmospheric windows, we adopt a fiducial
excitation temperature of $T_{\rm ex}=20$~K, consistent with the
average of three of the GEMS. Similar temperatures are found for
the lower-excitation component traced by CO lines by \citet{yang17}
and \citetalias{canameras18a}. By using the lowest representative
temperature measurement, we bias the optical depth of the lines high,
since the gas becomes optically thicker with decreasing temperature.
Had we adopted $T_{\rm ex}=37$ K instead (the highest excitation
temperature measured amongst the GEMS), we would have obtained optical
depths that are about 80\% lower.

\subsection{Mass of atomic carbon and carbon abundances}
\label{ssec:carbonabundance}

A major advantage of using optically thin lines for mass estimates is
that the line luminosity is proportional to the mass of the tracer.
We follow \citet{walter11} and \citet{weiss03} in estimating the mass
of atomic carbon by setting
\begin{equation}
{\rm M}_{\rm CI} = 5.71\times 10^{-4}\ Q(T_{\rm ex})\ 1/5e^{T_1/T_{ex}}\ L^\prime_{\rm CI10}\ [{\rm M}_{\odot}],
\end{equation}
where $T_{\rm ex}$ is the excitation temperature, and $Q(T_{\rm ex})$
the partition function $Q(T_{\rm ex}) = 1.0+3e^{-T_1/T_{\rm
ex}}+5e^{-T_2/T_{\rm ex}}$. $L^\prime_{\rm
CI10}$ are the measured luminosities of [CI] 1--0.

We use the measured excitation temperature, $T_{\rm ex}$, for the four
galaxies where we observed both [CI] line fluxes
(Table~\ref{tab:text}). The quantities $T_1=23.6$~K and $T_2=62.5$~K
correspond to the energies above the ground state for [CI] 1--0 and
[CI] 2--1, respectively. Results are listed in
Table~\ref{tab:masses}. Overall, we find that atomic carbon masses
are between $8\times 10^6$~M$_{\odot}$ and $5\times 10^7$
M$_{\odot}$ after correcting for the gravitational magnification
given in Table~\ref{tab:intrinsic}.

In principle, both lines of [CI] can be used as mass
tracers. \citet{weiss03} give an equivalent equation to Eq. (6) for
[CI] 2--1. However, in practice, estimates based on [CI] 2--1 are much
more sensitive to the excitation temperature. Whereas the mass
estimate derived from [CI] 1--0 changes by only about 1\% for a
temperature range between 20 and 50~K, mass estimates from [CI] 2--1
change by more than a factor of~3. Since the excitation temperature in the
four galaxies with [CI] 1--0 and [CI] 2--1 measurements does not
correlate with the dust temperature, to estimate robust excitation
temperatures from [CI] 2--1, we would need to observe both lines, in
which case we would estimate the atomic carbon mass directly from [CI]
1--0. We therefore do not derive carbon mass estimates for the GEMS
that have only [CI] 2--1 measurements.

Combining our mass estimates of atomic carbon with the molecular gas
mass estimates derived from CO by \citetalias{canameras18a} allows us
to estimate a carbon abundance, $X_{\rm CI} = X[CI]/X[H_2] = {\rm M}_{\rm C}
/ 6 {\rm M}_{\rm H_2}$. Obviously, a CO mass estimate must be chosen for
this calculation that does itself not depend on carbon abundance. We
are using the ULIRG conversion factor, $\alpha_{\rm CO, ULIRG}=\ 0.8\
{\rm M}_{\odot}$ / [K km s$^{-1}$ pc$^2$], which satisfies this
criterion. \citet{solomon97} derived this value from dynamical mass
estimates of nearby ULIRGs, supposing that the molecular gas mass
equals the difference between dynamical and stellar mass. For the same
reason, $\alpha_{\rm CO,ULIRG}$ also naturally accounts for He.

Using the total molecular gas mass estimates of
\citetalias{canameras18a}, and assuming, for the sake of this specific
analysis, that $\alpha_{\rm CO,ULIRG}$ is the perfect choice for these
targets (we will discuss this choice more broadly in the next
section), we find carbon abundances between 2.3 and $4.0\times$
10$^{-5}$ (Table~\ref{tab:masses}).

These abundance estimates are consistent with the canonical value
proposed by \citet{weiss05}, and derived for M82. They are also
consistent with previous work by \citet{alaghband13} and
\citet{danielson11}, for gravitationally lensed, dusty starburst
galaxies at similar redshifts; these authors found values between 3
and $4\times 10^{-5}$, comparable to what we find here. Several recent
analyses, however, come to different conclusions. For example,
\citet{bothwell17} found a high average carbon abundance of $X_{\rm
CI}=7.3\times 10^{-5}$ in a sample of 13 strongly lensed dusty
starburst galaxies from the SPT survey at $z\sim4$, when adopting
$\alpha_{\rm CO,ULIRG}$, whereas \citet{valentino18} very recently
found significantly lower values in a sample of main-sequence disk
galaxies at $z=1.2$, adopting a higher CO-to-H$_2$ conversion factor,
which is presumably more appropriate for main-sequence galaxies. We
will continue the discussion of the carbon abundances after deriving
molecular gas masses from the [CI] 1--0 luminosities in the next section.

\section{Molecular gas mass estimates from [CI] and CO}
\label{sec:gasmasses}

\begin{table*}
\caption{Diagnostic line ratios. \label{tab:ratios}}
\begin{center}
\begin{tabular}{lcccc}
\hline
\hline
Source        & CO transition&  $\mu$L$^\prime_{\rm CO}$\tablefootmark{a} &  $\mu$L$^\prime_{\rm CI10}$ & $L^\prime_{\rm CO}/L^\prime_{CI10}$ \\
              &           & [$10^{11}$  K km s$^{-1}$ pc$^2$] & [$10^{11}$ K km s$^{-1}$ pc$^2$] & \\
\hline
PLCK$\_$G045.1$+$61.1 & 4-3 & 7.5$\pm$0.9  & 2.3$\pm$0.5 & 3.3$\pm$0.8  \\
PLCK$\_$G080.2$+$49.8 & 3-2 & 2.9$\pm$0.2  & 1.1$\pm$0.2 & 2.6$\pm$0.5 \\
PLCK$\_$G092.5$+$52.9 & 4-3 & 10.9$\pm$0.7 & 3.3$\pm$0.6 & 3.3$\pm$0.6  \\
PLCK$\_$G102.1$+$53.6 & 3-2 & 2.2$\pm$0.8  & 0.7\tablefootmark{b} & 3.1 \\
PLCK$\_$G113.7$+$61.0 & 4-3 & 3.7$\pm$0.3  & 1.5$\pm$0.2 & 2.5$\pm$0.4  \\ 
PLCK$\_$G138.6$+$62.0 & 4-3 & 4.9 $\pm$0.3 & 1.5$\pm$0.2 & 3.3$\pm$0.5  \\
PLCK$\_$G145.2$+$50.9 & 4-3 & 12.2$\pm$2.4 & 4.2$\pm$1.0 & 2.9$\pm$0.9  \\
PLCK$\_$G165.7$+$67.0 & 4-3 & 4.6$\pm$0.3  & 1.5$\pm$0.4 & 3.1$\pm$0.8  \\
PLCK$\_$G200.6$+$46.1 & 3-2 & 6.0$\pm$0.6  & 1.3\tablefootmark{b} & 4.6 \\
PLCK$\_$G231.3$+$72.2 & 3-2 & 5.5$\pm$0.8  & 0.9\tablefootmark{b} & 6.4 \\
PLCK$\_$G244.8$+$54.9 & 4-3 & 7.0$\pm$0.7  & 3.1\tablefootmark{b} & 2.3 \\
\hline
\hline
\end{tabular}
\end{center}
\tablefoottext{a}{Taken from \citetalias{canameras18a}.}\\
\tablefoottext{b}{Estimated from [CI] 2--1, assuming a ratio
$I_{\rm CI1-0}/I_{\rm CI2-1}= 1.8$, the average of the values of the four
galaxies where we cover both lines. Error bars include the measurement
uncertainties, and are only given for galaxies where [CI] 1--0 was
actually measured. }\\
\end{table*}

\citet{weiss03}, \citet[][]{papadopoulos04a}, and
\citet[][]{wagg06} were among the first to propose the use of [CI] emission-line
measurements to estimate total molecular gas masses for high-redshift
galaxies. The main motivation was that these lines are bright and
optically thin, and that for a given carbon abundance and excitation
parameter, $Q_{10}$, a simple scaling between [CI] 1--0 line flux and
total gas mass can be given, as follows:

\begin{equation}
{\rm M}_{\rm H_2,[CI]} = 1380 \times \frac{D_{\rm L}^2}{(1+z)}\ A_{10,-7}^{-1}\ X_{{\rm CI},-5}^{-1}\ Q_{10}^{-1}\ I_{\rm CI}\ [{\rm M_{\odot}}],
\end{equation}
$D_{\rm L}$ is the luminosity distance in units of Gpc, $z$ the
redshift, $I_{\rm CI}$ the integrated line flux of [CI] 1--0 in Jy km
s$^{-1}$. The Einstein A coefficient for [CI] 1--0, $A_{10}$ is given
in units of $10^{-7}$ s$^{-1}$, and the carbon abundance, $X_{\rm CI}$, is
in units of $10^{-5}$.

We set $X_{\rm CI}=3\times 10^{-5}$, the standard value that has also been
commonly adopted in previous work \citep[e.g.,][]{walter11}, and
$A_{10}=7.93\times 10^{-8}$, similar to
previous authors. For $Q_{10}$ we adopted 0.49, the median $Q_{10}$
value used by \citet{papadopoulos04a} and also used previously by
\citet{alaghband13}. 

With the flux measurements listed in Table~\ref{tab:lineproperties},
between 3 and 21~Jy km s$^{-1}$, and the redshifts listed in the same
table, we find total molecular gas mass estimates, ${\rm M}_{\rm
H2,CI}$ between 10 and $40\times 10^{11}\ \mu^{-1}$ M$_{\odot}$ for
the seven galaxies that have [CI] 1--0 measured. Results for
individual sources are listed in Table~\ref{tab:masses}.

In Table~\ref{tab:masses}, we also compare with molecular gas mass
estimates derived from CO line emission, for a CO-to-H$_2$ conversion
factor of $\alpha_{\rm CO,ULIRG} = 0.8\ {\rm M}_{\odot}$/ [K km
s$^{-1}$ pc$^2$]. We follow \citetalias{canameras18a},
who derived gas masses from the measured CO(4--3) or CO(3--2)
luminosities, depending on which line falls into the atmospheric
windows, and taking CO(3--2) when both lines are available. These
luminosities, L$^\prime$, were corrected by ratios of
$R_{32}=L^\prime_{\rm CO(3-2)}/L^\prime_{\rm CO(1-0)}=0.4$ and
$R_{43}=L^\prime_{\rm CO(4-3)}/L^\prime_{\rm CO(1-0)}=0.3$
\citepalias{canameras18a} to extrapolate to $L^\prime_{\rm
CO(1-0)}$. \citetalias{canameras18a} derived these average line ratios
by comparing with the CO(1--0) mass estimates of \citet{harrington18},
which are available for four GEMS, for $\alpha_{\rm CO,ULIRG} = 0.8\
{\rm M}_{\odot}$/ [K km s$^{-1}$ pc$^2$]. We will in the following use
these ratios to adopt a common procedure for our entire sample,
including galaxies with and without CO(1--0) measurements.

With these line ratios, we find an excellent agreement between the
masses derived from [CI] and CO for all GEMS that have [CI] 1--0
observations. Amongst the four sources with CO(1--0) measurements, three
 have consistent mass estimates from CO(1--0) and [CI]
1--0 within the measurement uncertainties. Only one source,
PLCK$\_$113.7$+$61.1 has a somewhat higher mass estimate from [CI]
1--0 than from CO(1--0), with a ratio
M$_{H2,CI10}$/M$_{H2,CO10}=1.4\pm0.1$. For the overall sample, and
using molecular gas mass estimates derived from CO(4--3) or CO(3--2),
five of seven sources have consistent mass estimates (within 2$\sigma$),
and two sources have somewhat larger mass estimates from [CI] 1--0
than from CO, M$_{H2,CI10}$/M$_{H2,CO10}=1.7\pm0.2$. Individual
results are listed in Table~\ref{tab:masses}.

Finding consistent mass estimates with two independent tracers is
certainly an encouraging result, and may serve as a validation of
applying low-redshift calibrations to (at least this type of)
high-redshift galaxies. It confirms that using low, average
CO-to-H$_2$ conversion factors akin to the used factor $\alpha_{\rm
CO,ULIRG}=0.8\ {\rm M}_{\odot}$ [K km s$^{-1}$ pc$^2$]
\citep[][]{solomon97}, average carbon abundance of about
$X_{\rm CI}=3\times 10^{-5}$ \citep[][]{weiss05}, and excitation parameter
$Q_{10}\sim 0.5$ \citep[][]{papadopoulos04a} is an adequate,
internally consistent choice of parameters.

However, this result should be interpreted with some caution. In
particular, it cannot be used as a justification for any peculiar
choice of $X_{\rm CI}$ or $\alpha_{\rm CO}$, since both are degenerate, as
inserting the expression of carbon abundance explicitly into Eq.~(7)
immediately shows. Consequently the largest systematic uncertainty in
carbon abundance measurements, and in molecular gas mass estimates
from [CI] is still $\alpha_{\rm CO}$.

The total atomic carbon mass estimates, however, are independent of
the chosen $\alpha_{\rm CO}$. Therefore, significantly increasing
$\alpha_{\rm CO}$ for galaxies like the GEMS would also imply that we should adopt 
equally low carbon abundances. We argued in C15 that the
metallicities in the GEMS are proabably already solar or greater,
relying on gas-to-dust ratios as previously estimated by
\citet{magdis11}. Solar or greater gas-phase metallicities in massive,
dusty starburst galaxies at redshifts $z\sim 2 - 3$ are also suggested
by studies of warm ionized gas in these galaxies
\citep[][]{takata06,nesvadba07}, as well as by the abundances found in
the photospheres of the dominant stellar populations in massive
low-redshift galaxies, which probe the metallicities at the time when
these stars were formed;  they are solar or super-solar
\citep{gallazzi05}, and also do not favor unusually low carbon
abundances in the GEMS and other, similar high-redshift galaxies. This
makes a much lower $X_{CI}$, and a higher $\alpha_{\rm CO}$ than the
ULIRG-value implausible, at least for this type of high-redshift
galaxy. For bluer, lower-mass, and less intensely star-forming
galaxies, this is probably different, and overall, the range of
$\alpha_{\rm CO}$ is probably set by a range of parameters, including in
particular metallicity \citep[e.g.][]{bolatto13}.
 
\begin{figure}
\includegraphics[width=0.45\textwidth]{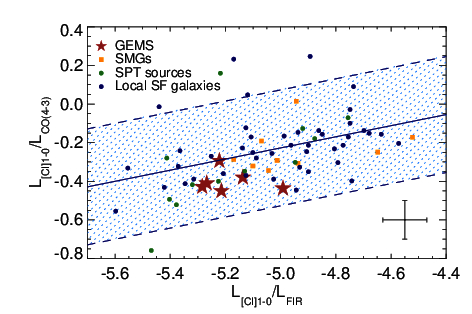}
\caption{\label{fig:cicoratio} Ratio of $L_{\rm CI1-0}$ to $L_{CO(4-3)}$ as
a function of the cooling budget through [CI] 1--0. The red stars are
the GEMS. Blue, green, and orange dots indicate the samples of
low-redshift star-forming galaxies from \citet{kamenetzky16}, and the
high-redshift samples of gravitationally lensed sources from the SPT
\citep{bothwell17} and the submillimeter galaxies of
\citet{alaghband13}, respectively. The blue line shows the average
relationship derived from the comparison samples. The blue hatched
region shows a range of a factor of~$\pm2$ around this average. The typical
measurement uncertainty is shown in the lower right corner.}
\end{figure}

Collecting large enough sets of emission lines of individual
high-redshift galaxies to study their gas excitation in detail is
often very challenging. To obtain these mass estimates, it was
critical in our case to accurately take into account gas excitation
when extrapolating from mid-$J$ CO line luminosities to the
luminosities of CO(1--0), since the line ratios are lower by factors
$1.5-2$ than others in the literature \citepalias[for details
see][]{canameras18a}. Had we used the higher values of, e.g.,
\citet[][]{spilker14} or \citet[][]{danielson11}, we would have been
led to conclude to have found considerably higher molecular gas masses
from [CI] 1--0 than from the mid-$J$ CO
lines. Figures~\ref{fig:cicooling} and~\ref{fig:cicoratio} show that
the GEMS as an ensemble have somewhat lower ratios of $L_{\rm
CI10}/L_{CO43}$, and somewhat higher ratios of $L_{\rm CI10}/L_{\rm
FIR}$ compared to other populations of high and low-redshift galaxies,
which is consistent with this finding. This may indicate that their
gas is perhaps somewhat denser or more highly excited than in other
galaxies at the same redshifts, as also shown by
\citetalias{canameras18a}, and as would be consistent with targeting
particularly bright galaxies on the sub-millimeter sky. Despite these
indications, they fall well within the scatter of the overall
population.

If our results are applicable to more general populations of massive,
dusty, high-redshift starburst galaxies, then this would imply that
most of the discrepancy seen in mass estimates from [CI] and CO could
be dominated by the diversity in average gas excitation of these
galaxies. For example, similar effects could be at play for other
samples of vigorous starburst galaxies showing enhanced carbon
abundances, like those found with the SPT \citep[][]{bothwell17}. The
origin of this diversity might either be differences in the excitation
process itself, or in the relative contribution of high and
low-excitation gas \citep[e.g.,][, C18]{ivison10, harris10, yang17}.
Even the largest ratios of ${\rm M}_{H2,CI}$ to ${\rm M}_{H2,CO}$
amongst the GEMS, namely ${\rm M}_{H2,CI}/ {\rm M}_{H2,CO} =
1.7\pm0.2$, could reflect differences in gas excitation rather than
additional gas components not seen in CO(1--0). For example,
\citet{papadopoulos04b} state a range of a factor of~3 of plausible
excitation parameters $Q_{10}$ for molecular gas mass estimates from
[CI] 1--0.

Regardless of these concerns, our results do suggest that CO(1--0) and
[CI] 1--0 are probing similar gas reservoirs within the GEMS, and that
the impact of differential lensing does not dominate the observed
luminosity and mass estimates derived from either line. In particular,
and while we do see multiple gas components with different excitation
conditions in the GEMS \citepalias{canameras18a}, we find no evidence
that such galaxies have large fractions of ``CO-dark'' cold,
neutral gas, that would not be seen in CO(1--0).

\begin{table*}
\caption{ Molecular gas mass estimates derived from [CI] 1-0, and
from CO. Ratios of mass estimates and Carbon abundance, $X_{\rm CI}$, for
the different molecular gas mass estimates from CO. \label{tab:masses}}
\begin{center}
\begin{tabular}{lcccccc}
\hline
\hline
  Source         & $\mu\ {\rm M}_{\rm H_2,CI}$\tablefootmark{a} & $\mu\ {\rm M}_{\rm H_2,CO43,extr}$\tablefootmark{b}&  ${\rm M}_{\rm H2,CI}/{\rm M}_{\rm H2,CO10}$\tablefootmark{c} &  ${\rm M}_{\rm CI}/{\rm M}_{\rm H2,CO43,extr}$ & $X_{\rm CI,CO10}$ & $X_{\rm CI,CO43,extr}$ \\
                 &[$10^{11} {\rm M}_{\odot}$]  &  [$10^{11} {\rm M}_{\odot}$] &     &   &  [$\times 10^{-5}$] & [$\times 10^{-5}$]\\ 
\hline 
PLCK$\_$G045.1$+$61.1  & 20.9$\pm$4.7 & 19.9$\pm$2.4 &   \dots       & 1.0$\pm$0.3    & \dots       & 2.5$\pm$0.5   \\
PLCK$\_$G080.2$+$49.8  & 9.8$\pm$0.8  & 5.7$\pm$0.4  &   \dots       & 1.7$\pm$0.2    & \dots       & 3.8$\pm$0.9   \\
PLCK$\_$G092.5$+$52.9  & 30.9$\pm$2.3 & 24.8$\pm$1.6 &  1.1$\pm$0.4  & 1.2$\pm$0.1    & 2.7$\pm$0.9 & 3.0$\pm$0.3   \\
PLCK$\_$G113.7$+$61.0  & 14.2$\pm$0.7 & 11.6$\pm$0.9 &  1.4$\pm$0.1  & 1.2$\pm$0.1    & 3.2$\pm$1.0 & 2.9$\pm$0.6   \\
PLCK$\_$G138.6$+$62.0  & 13.4$\pm$0.9 & 14.6$\pm$0.9 &  1.1$\pm$0.4  & 0.9$\pm$0.1    & 2.6$\pm$0.7 & 2.2$\pm$0.4   \\
PLCK$\_$G145.2$+$50.9  & 38.9$\pm$0.5 & 23.3$\pm$0.9 &  \dots        & 1.7$\pm$0.1    & \dots       & 4.0$\pm$0.7   \\
PLCK$\_$G165.7$+$67.0  & 13.5$\pm$0.4 & 15.0$\pm$0.4 &  0.95$\pm$0.1 & 0.9$\pm$0.1    & 2.3$\pm$0.5 & 2.2$\pm$0.3   \\
PLCK$\_$G244.8$+$54.9  &  \dots       & 14.0$\pm$1.4 &   \dots       & \dots          &  \dots      &   \dots       \\
\hline
\hline
\end{tabular}
\end{center}
\tablefoottext{a}{For $X_{\rm CI}=3\times 10^{-5}$, and $\alpha_{\rm CO,ULIRG}=0.8\ {\rm M}_{\odot}$ / [K km s$^{-1}$ pc$^2$].}\\
\tablefoottext{b}{Using the average luminosity ratios of the GEMS, $R_{32}=L^\prime_{\rm CO(3-2)}/L^\prime_{\rm CO(1-0)}=0.4$ and $L^\prime_{\rm CO(4-3)}/L^\prime_{\rm CO(1-0)}=0.3$ \citepalias{canameras18a} to extrapolate to $L^\prime_{\rm CO(1-0)}$.}\\
\tablefoottext{c}{CO(1--0) is taken from Harrington et al. (2018), for $\alpha_{\rm CO,ULIRG}=0.8\ {\rm M}_{\odot}$ / [K km s$^{-1}$ pc$^2$].}
\end{table*}

\section{Summary and conclusions}
\label{sec:summary}

We have presented an analysis of the [CI] 1--0 and [CI] 2--1 emission
lines in {\it Planck}'s Dusty GEMS, a small sample of 11 of the
brightest high-redshift galaxies on the sub-millimeter sky observed
with the {\it Planck} satellite. We have detailed lens models derived
with {\sc Lenstool} from sub-arcsecond interferometry for all galaxies
\citep{canameras18a}, and can therefore explicitly account for source
morphology and differential lensing between dense gas and dust
(finding that it does not play a major role). We detect all [CI] lines
from those galaxies where these lines fall into the atmospheric
windows. In total, we measured the [CI] 1--0 line in seven, and the
[CI] 2--1 line in eight galaxies, with four galaxies having
measurements of both lines. Our main results are as follows.

\begin{itemize}
\item
  The GEMS have [CI] line fluxes between 4 and 21~Jy~km~s$^{-1}$, with
$L_{\rm [CI]21+10}/L_{\rm FIR}$ between $1.2\times 10^{-5}$ and $2.6\times 10^{-5}$,
comparable, and in the lower range of other galaxies at low and high
redshift.
\item
  Line ratios $L_{\rm CI21}/L_{\rm CI10}$ are between 1.2 and 3.3, and
the [CI] line emission is consistent with optically thin
{\bf ($\tau=0.01-0.14$)} gas in star-forming regions dominated by UV
heating, without major contribution from an AGN, and excitation
temperatures of typically about $T_{\rm ex}=20$~K, with one galaxy
having T$_{\rm ex}=36$~K.
\item
  The line ratios of [CI] 1--0 and CO(4-3) are between 2.3 and
3.5. Following \citet{greve12} and \citet{papadopoulos12} we interpret
this as a sign that these galaxies are undergoing starbursts, not
the more regular, longer-term star formation typical of disk galaxies
at similar redshifts.
\item
  The intrinsic masses of atomic carbon are beween 0.8 and $6.3\times
10^7$~M$_{\odot}$, corresponding to atomic carbon abundances between 
$X_{\rm CI}= 2\times 10^{-5}$ and $4\times 10^{-5}$. This is comparable to 
the usually adopted value of $3\times 10^{-5}$ initially derived for
M82, and several other samples of high-redshift galaxies. However,
recent studies have also found either higher \citep[][]{bothwell17} or
lower values \citep[][]{valentino18}, in either case within a factor
of about~2.
\item
 ${\rm H_2}$ gas mass estimates from [CI] 1--0 (and adopting a
carbon abundance of $3\times 10^{-5}$), correspond to those measured
from CO within the measurement uncertainties for five of seven
galaxies that have [CI] 1--0 measured, and within factors 1.7 for the
other two. These values were derived for a standard ''ULIRG''
CO-to-H$_2$ conversion factor, $\alpha_{\rm CO}=0.8\ {\rm M}_{\odot}$/ [K km
s$^{-1}$ pc$^2$], and from mid-J CO line observations (either $J$=4--3 or
$J$=3--2) corrected for their ratio with CO(1--0), as directly observed
by \citet{harrington18} for four GEMS. These ratios are factors
of $1.5-2$ lower than previously proposed for other samples of massive,
dusty starburst galaxies at comparable redshifts
\citep[][]{canameras18a}, suggesting that the gas excitation
conditions play a non-negligible role in molecular gas mass estimates
of dusty starburst galaxies at redshifts $2-4$. Once
excitation was properly taken into account, we found that the standard
values of $\alpha_{\rm CO}$, atomic carbon abundances, and [CI] excitation
parameter $Q_{10}=0.49$, together give consistent results for molecular
gas mass estimates derived from [CI] and CO in these
galaxies. Consequently, we do not see evidence for large gas
reservoirs that are only probed by [CI] but not CO.
\end{itemize}

\section*{Acknowledgments}
We would like to thank the anonymous referee for interesting comments
which helped significantly improve the paper.  We also thank the
telescope staff at the IRAM~30-m telescope for their excellent support
during observations. We are particularly grateful to the former
director of IRAM, P. Cox, for his generous attribution of Director's
Discretionary Time early on during the program, and to C.~Kramer for
sharing his {\tt FtsPlatformingCorrection5.class} with us, which
significantly helped improve the quality of our spectra. RC was
supported by DFF-- 4090-00079, and CY by an ESO fellowship. This work
was supported by the Programme National Cosmologie et Galaxies (PNCG)
of CNRS/INSU with INP and IN2P3, co-funded by CEA and CNES. IRAM is
supported by INSU/CNRS (France), MPG (Germany), and IGN (Spain).
\bibliographystyle{aa}
\bibliography{lens}

\end{document}